# Iterative Methods for Systems' Solving - a C# approach

**Claudiu Chirilov**
"Tibiscus" University of Timişoara, România

ABSTRACT. This work wishes to support various mathematical issues concerning the iterative methods with the help of new programming languages. We consider a way to show how problems in math have an answer by using different academic resources and different thoughts. Here we treat methods like Gauss-Seidel's, Cramer's and Gauss-Jordan's.

## Introduction

Nowadays, it seems that a developer's problem is not to solve mathematical issues rather than finding solution for the problems in real life. For that reason, many of us do not know that new technologies coming into our life will not just solve things like how to keep track of events or how to improve a website's design. Therefore, in the end all problems have a mathematical background. This article will treat some mathematical stuff, such as iterative methods using a new programming language.

As we all should know, C# is very flexible programming language that combines features from previous languages in such manner that it is so easy to digest and fun to work with. C# is really a better playground from those coming from other programming languages. It is the base of the .NET Platform.

## 1. Mathematical background

In this article we will study and to analyze how some of the classical iterative methods can be implemented in C# for solving systems of linear equations, like the Cramer's method, Gauss-Jordan or Gauss-Seidell's method.

An iterative method, in computational mathematics, is about the attempt to solve a problem by finding succesive approximations to the solution starting from an initial guess. This type of solving refers not just to nonlinear equations but it also applies to linear systems of equations.

A system of linear equations, in mathematics, is a collection of linear equations in wich a set of variables verifies all the equations at the end. Let's say we have the folowing system of linear equations:

$$\begin{cases} a_{11}x_1 + a_{12}x_2 + \ldots + a_{1n}x_n = b_1 \\ a_{21}x_1 + a_{22}x_2 + \ldots + a_{2n}x_n = b_2 \\ \ldots \\ a_{m1}x_1 + a_{m2}x_2 + \ldots + a_{mn}x_n = b_m \end{cases}$$

The solution of this system must satisfy in the same time all the equations. We want to solve the system by applying different methods and algoritms for finding the solution.

Cramer's method is based on the determinants of the system matrix and transformed matrices. Here is how the method works:
   a) Compute the determinant of the system matrix.
   b) Build the transformed matrices by replacing each column of the system matrix with the results matrix.
   c) Now for each transformed matrix divide its determinant by the determinant of the system matrix.

The Gauss-Jordan's method has more steps to follow and performs elementary operations on the rows of the matrix. The algorithm says:
   a) If they are zero-rows, push them to the bottom of the matrix, by succesive row interchanges.
   b) Find the first column containing at least one non zero entry.
   c) Eventually interchange two rows so that this non zero entry will be the first in this column.
   d) If there is another row having a non-zero entry in this column, say $R_i$, perform the folowing operation: $R_i \leftarrow R_i - \frac{a_{ik}}{a_{1k}} R_1$. This replaces $a_{ik}$ by 0.
   e) Iterate this step until you replace all entries below $a_{1k}$ by zeros.

f) If the matrix is still not in row-echelon form, move to the next column having a non zero entry, below the level of the leading entry you worked with and perform similar operations.

g) After you got a row-echelon form for your matrix, perform elementary operations to put the leading entries to 1 and to force the entries above them to zeros.

The Gauss-Seidel iterative method represents an improvement of the Jacobi iteration and it is also called the method of succesive displacements. Briefly, the algorithm consists in the following steps:

- Verification of the coefficient matrix values, the method being used only for matrix with nonzero diagonal entries
- Rewrite each equation solving for the corresponding unknown
- Use rewritten equations to solve for each value of $x_i$
- The iterations are stopped when the absolute relative approximate error is less than an early-specified tolerance for all unknowns.

## 2.C# (Theoretical aspects)

For our purpose we choosed one external library in order to help us doing some matrix operations. The **CSML (CSharp Matrix Library)** has the tool we need to do operations like the inverse of a matrix, calculate the determinant and so on. As we know all programming languages must have a common math functions library. C# has it with no exception. You can access the mathematical functions in C# by importing the namespace **System.Math** at the beginning of the source code: **using System.Math;** .

Most of the variables defined in the source are single precision floating-point type (**System.Single** ) and integer-type variables ( **System.Int32** ). The scope was to represent the results as accurate as we can. We choose to create a simple console application for our mean.

We used the Academic Resource Kitt tools to solve these methods in C#.

## 3.Inside the code

Another interesting method in wich the use of **CSML (Csharp Matrix Library)** is imperative is the **Cramer method**. It helps solving linear

73

systems of equations. Before starting to see how the code looks, it's necessary to add a reference to this library, to import it in your project. You can do this by adding this line to the beginning of your source file: **using CSML;** . The constructor of the **Matrix** class in the **CSML** library is overloaded and we opt for the one in wich we pass a string matrix. The library enforces us to format the input of the matrix values into the folowing format: **x, y, z ; w, q, p ; . . .** . In our source code we do this by calling two function, one for the **coefficients matrix** and one for the **results matrix**:

```
string strMatriceA = MatriceaCoeficientilorSistemului(a, n);
string strMatriceB = MatriceRezultatelorSistemului(b, n);
```

With the string matrix, strMatriceA can be formed a new **Matrix** object to calculate its determinant.

```
Matrix matriceA = new Matrix(strMatriceA);
```

Then we call a function to form as much transformed matrices as the rank of the coefficients matrix is. For each of these transformed matrices we obtain matrix strings in order to instantiate a new **Matrix** object.

```
string[] strMBuilder = new string[n];
for (int i = 0; i < n; i++)
{
     strMBuilder[i] = matriceaTransformata(b, a, i);
}

Matrix[] matriceTransformata = new Matrix[n];
for (int i = 0; i < n; i++)
{
    matriceTransformata[i] = new Matrix(strMBuilder[i]);
}
```

Each transformed matrix is actually the coefficients matrix but every column it's replaced by the results matrix one by one. The solve of the system finishes when we compute the result of the division between the determinat of each transformed matrix and the determinant of the coefficients matrix.

```
Complex[] rezultat = new Complex[n];
for (int i = 0; i < n; i++)
{
```

```
rezultat[i] = solutie(matriceTransformata[i].Determinant(),
matriceA.Determinant());
}
```

In the end, all we should do is to print the results on the screen using the **Console.WriteLine()** method:

```
for (int i=0;i<n;i++)
    Console.WriteLine("Solutia x{0} este: " +
rezultat[i].ToString(),i);
```

This method solves the systems of linear equations but it's rather an exact method than an iterative one. The results are exact and it doesn't shows us an approximative outcome. It is a good choice in order to obtain, if possible, an exact solution of the system, solution that can be later used as reference value for verifying iterative methods.

For the other methods that we will reffer to things are a bit different. This is the case for the **Gauss-Seidel** method. The **Gauss-Seidel** method is sometimes called the *method of succesive displacements* to indicate the dependence of the iterates on the ordering. If this ordering is changed, the *components* of the new iterate (and not just their order) will also change.

The method accepts the initial guess of the solutions. Starting from these ones the program will display approximations to the solutions. We also need an epsilon variabile to display the solution more accuratelly.

```
private static double epsilon = 0.01;
```

For the first part of the program we don't need nothing else to do just to read the augmented matrix and the system results matrix. Then we will have to enter the guess for the solutions. Before going to the main part we test the system matrix to see if the diagonal elements are 0, otherwise the solutions are not possible to compute without pivoting.

```
for(i=0;i<n;i++)
{
  if(a[i,i]==0)
      Console.WriteLine("Elementele de pe diagonala
        principala sunt 0,      nu exista solutii fara
        pivotare.\n");
}
```

75

The main part of the program that solves the system has the following body:

```
            for(i = 0;i < n;i++)
            {
                s = 0;
                for (j = 0; j < n; j++)
                {
                    if (i != j)
                        s += a[i, j] * x[j];
                }

                Xi = (b[i]-s)/a[i,i];

                Console.WriteLine("\nx[{0}]={1}",i,xi);

                d = System.Math.Abs(xi-x[i]);

                Console.WriteLine("d={0}",d);

                x[i] = xi;

                Console.WriteLine("x[{0}]={1}",i,x[i]);
            }
        }
        while ( d >= epsilon);
```

Finally we print the solution that is stored in the initial guess array, x.:

```
Console.WriteLine("\nRezultate sistem:");
for (i = 0; i < n; i++)
{
    Console.WriteLine("\n\tx[{0}]={1} ", i, x[i]);
}
```

The methods above were tested on a particular system, wich has .NET® Framework 2.0 installed.

**Conclusions**

The above techniques were made on a particular system of linear equations:

$$2x + 3y - z = 5$$
$$4x + 4y - 3z = 3 \ ,$$
$$-2x + 3y - z = 1$$

and the solution was $\begin{array}{l} x = 1 \\ y = 2 \\ z = 3 \end{array}$.

The Gauss-Jordan elimination method is one of the quickest, easy understandable and flexible for implementation in any programming language. It has much simpler rules than the more familiar techniques following algebraic substitution or the use of determinants (the Cramer's method). However, the method has a major disadvantage: it is up to three times slower than other systems' solving techniques and it requires additional storage resources.

The Gauss-Seidel iteration converges much faster than other methods and the rate of convergence is growing if the coefficient matrix is strictly diagonal dominant. Its main advantage is that the value computed by the Gauss-Seidel iteration is immediately used for improvement in other components. As presented in [CB81] the techique is called "sweep".

The methods presented in this paper are usually applied to large linear systems with sparse coefficient matrix. In this case, the number of operations at each iteration is reduced and the result is not affected by the round-off errors. Disadvantages appear when we have to use non-sparse matrix, because iterative methods do not always converge or require a large number of iterations.